\documentclass{PoS} \usepackage{epsfig}

\title{SU(3) Deconfining Phase Transition in a Box with Cold Boundaries}

\ShortTitle{SU(3) Deconfining Phase Transition in a Box with Cold Boundaries}

\author{Alexei Bazavov\thanks{Present address: University of Arizona,
Department of Physics, Tucson, AZ 85721.}$\ $ and \speaker{Bernd Berg}%
\\ Florida State University, Department of Physics, 
   Tallahassee, FL 32306-4350, USA\\ and
\\ Florida State University, School of Computational Science,
   Tallahassee, FL 32306-4120, USA\\
    E-mail: \email{bazavov@scs.fsu.edu}, \email{berg@hep.fsu.edu}}

\abstract{Deconfined regions created in heavy ion collisions are 
bordered by the confined phase. We discuss boundary conditions (BCs) 
to model a cold exterior. Monte Carlo simulations of pure SU(3) lattice 
gauge theory with thus inspired BCs show scaling. Corrections to usual 
results survive in the finite volume continuum limit and we estimate 
them in a range from $L=5-10$ fermi as function of the volume size 
$L^3$. In magnitude these corrections are comparable to those obtained
by including quarks. }

\FullConference{The XXV International Symposium on Lattice Field Theory\\
                 July 30 - August 4 2007\\
                 Regensburg, Germany}

\begin{document}

\section{Introduction}

Past LGT simulations of the deconfining transition focused primarily 
on boundary conditions (BCs), which are favorable for reaching the 
infinite volume quantum continuum limit (thermodynamic limit of the 
textbooks) quickly.  On $N_{\tau}\,N_s^3$ lattices these are periodic 
BCs in the spatial volume $V = (a\,N_s)^3$, where $a$ is the lattice 
spacing. The temperature of the system is given by
\begin{equation} 
  T = \frac{1}{a\,N_{\tau}} = \frac{1}{L_{\tau}}\,,~~~
  (N_{\tau}<N_s)\ .
\end{equation}
In the following we set the physical scale by 
\begin{equation} 
  T^c=174\ {\rm MeV}\ ,
\end{equation}
which is approximately the average from QCD estimates with two light 
flavor quarks. This implies for the temporal extension
\begin{equation} 
  L_{\tau} = a\,N_{\tau} = 1.13\ {\rm fermi}\ .
\end{equation}
For the deconfinement phase created in a heavy ion collision the 
infinite volume limit does not apply. Instead we have to take the 
finite volume continuum limit
\begin{equation}
  N_s/N_{\tau} = {\rm finite}\,,~~N_{\tau}\to\infty\,,~~L_{\tau}~~
  {\rm finite}\,,
\end{equation}
and periodic BCs are incorrect, because the outside is in the 
confined phase at low temperature. E.g., at the BNL RHIC one expects 
to create an {\it ensemble of differently shaped and sized deconfined 
volumes.} The largest volumes are those encountered in central 
collisions. A rough estimate of their size is
\begin{eqnarray} \nonumber
  & \pi\times (0.6\times {\rm Au\ radius})^2\times c\times
  ({\rm expansion\ time}) &  \\
  & = (55\ {\rm fermi}^2)\times ({\rm a\ few\ fermi}) &
\end{eqnarray}
where $c$ is the speed of light. Here we report on our work 
\cite{BaBe07}, which estimates such finite volume corrections 
for pure SU(3) and focuses on the continuum limit for
\begin{equation} 
  L_s = a N_s = (5-10)\ {\rm fermi}\ .
\end{equation}

\section{Equilibrium with Unconventional Boundary Conditions}

Statistical properties of a quantum system with Hamiltonian $H$ in
a continuum volume $V$, which is in equilibrium with a heatbath 
at physical temperature $T$, are determined by the partition 
function
\begin{equation}
    Z(T,V)={\rm Tr} e^{-H/T}=
    \sum_{\phi}\langle\phi|e^{-H/T}|\phi\rangle,
\end{equation}
where the sum extends over all states and the Boltzmann constant is 
set to one. Imposing periodic boundary conditions in Euclidean time 
$\tau$ and bounds of integration from $0$ to $1/T$, one can rewrite 
the partition function in the path integral representation:
\begin{equation}
    Z(T,V)=\int D\phi\exp\left\{-\int_0^{1/T}d\tau L_E
    (\phi,\dot\phi)\right\}.
\end{equation}
Nothing in this formulation requires to carry out the infinite
volume limit. 

In the following we consider difficulties and effects encountered when 
one equilibrates a hot volume with cold boundaries by means of Monte
Carlo (MC)
simulations for which the updating process provides the equilibrium. 
We use the single plaquette Wilson action on a 4D hypercubic lattice.  
Numerical evidence suggests that SU(3) lattice gauge theory exhibits 
a weakly first-order deconfining phase transition at some coupling 
$\beta^g_t(N_{\tau})=6/g^2_t (N_{\tau})$. The scaling behavior of the 
deconfining temperature is
\begin{equation}
  T^c = c_T\,\Lambda_L
\end{equation}
where the lambda lattice scale
\begin{equation}\label{LambdaLat} 
  a\,\Lambda_L = f_{\lambda}(\beta^g) = \lambda(g^2)\,
  \left(b_0\,g^2\right)^{-b_1/(2b_0^2)}\,e^{-1/(2b_0\,g^2)}\,,
\end{equation}
has been determined in the literature. The coefficients $b_0$ and $b_1$ 
are perturbatively determined by the renormalization group equation:
\begin{equation} 
  b_0 = \frac{11}{3}\frac{3}{16\pi^2}~~{\rm and}~~
  b_1=\frac{34}{3}\left(\frac{3}{16\pi^2}\right)^2\,.
\end{equation}
Relying on work by the Bielefeld group \cite{Bo96} we parametrized in
\cite{BBV06} higher perturbative and non-perturbative corrections by
\begin{equation} 
  \lambda(g^2)\ =\ 1+a_1\,e^{-a_2/g^2}+a_3\,g^2+a_4\,g^4
  ~~~{\rm with}
\end{equation}
$a_1=71553750,\ a_2=19.48099,\ a_3=-0.03772473,\ a_4=0.5089052$,
which turns out to be in good agreement with \cite{NeSo02} in
the coupling constant range for which the latter is claimed to
be valid.

Imagine an almost infinite space volume $V=L_s^3$, which may have 
periodic BCs, and a smaller (very large, but small compared to $V$)
sub-volume $V_0=L_{s,0}^3.$ The complement to $V_0$ in $V$ will 
be called $V_1$. The number of temporal lattice links $N_{\tau}$ is 
the same for both volumes.  We denote the coupling by $\beta^g_0$ for 
plaquettes in $V_0$ and by $\beta^g_1$ for plaquettes in $V_1$. For 
that purpose any plaquette touching a site in $V_1$ is considered to 
be in $V_1$. This defines a BC, which we call {\it disorder wall}.

We would like to find couplings so that scaling holds, while $V_0$ 
is at temperature $T_0=174\, {\rm MeV}$ and $V_1$ at room temperature 
$T_1$. Let us take $\beta^g_1=5.7$ at the beginning of the SU(3) 
scaling region. We have
\begin{equation}
  10^{10} \approx \frac{T_0}{T_1} = \frac{a_1}{a_0} =
  \frac{f_{\lambda}(\beta^g_1)}{f_{\lambda}(\beta^g_0)}
\end{equation}
where $a_i$ is the lattice spacing in $V_i,\, i=0,1$.  Using
the lambda scale yields $\beta^g_0\approx 25$ and $T_c$ 
estimates of the literature give $L_{\tau} > 10^{11}\,a$. 
In practice we can only have $\beta^g_0$ in the scaling region.
We keep up the relation
\begin{equation}
  \frac{(\xi/a_0)}{(\xi/a_1)} = \frac{a_1}{a_0} \approx 10^{10}
\end{equation}
where $\xi$ is a correlation length. Therefore, $\xi/a_1$ is 
very small and the strong coupling expansion implies 
$\beta^g_1\approx 10^{-10^{10}}$, i.e., $\beta^g_1=0$.

\section{Monte Carlo Calculations with Disorder Wall BCs} 

In the disorder wall approximation of the cold exterior we can simply 
omit contributions from plaquettes, which involve links through the 
boundary. Due to the use of the strong coupling limit for the BCs, 
scaling of the results is not obvious.

We use the maxima of the Polyakov loop susceptibility
\begin{equation}
  \chi_{\max}=\frac{1}{N_s^3}\left[\langle|P|^2\rangle-
  \langle|P|\rangle^2\right]_{\max},\,\,\,P=\sum_{\vec{x}}P_{\vec{x}}
\end{equation}
to define pseudo-transition couplings $\beta^g_{pt}(N_s;N_{\tau})$.
For periodic BCs they have a finite size behavior of the form
\begin{equation} 
  \beta^g_{pt}(N_s;N_{\tau}) = \beta^g_t(N_{\tau}) +
  a_3^p\,\left(\frac{N_{\tau}}{N_s}\right)^3 +\ \dots\ .
\end{equation}
Our BCs introduce an order $N_s^2$ disturbance, so that 
\begin{equation} 
  \beta^g_{pt}(N_s;N_{\tau}) = \beta^g_t(N_{\tau}) +
  a_1^d\,\frac{N_{\tau}}{N_s} +
  a_2^d\,\left(\frac{N_{\tau}}{N_s}\right)^2 +
  a_3^d\,\left(\frac{N_{\tau}}{N_s}\right)^3 +\ \dots\ .
\end{equation}
%
\begin{figure}[-t] \begin{center} 
\epsfig{figure=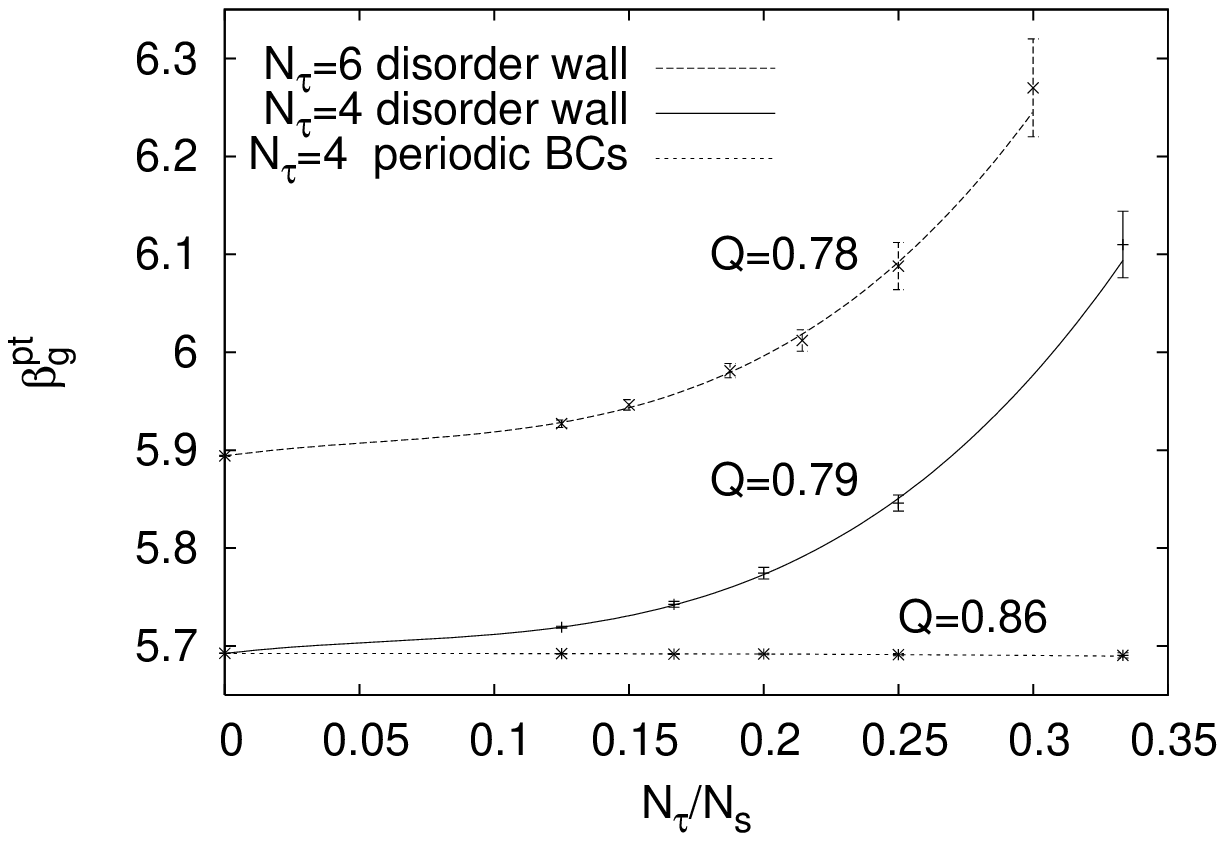,width=0.47\textwidth}
\hfill
\epsfig{figure=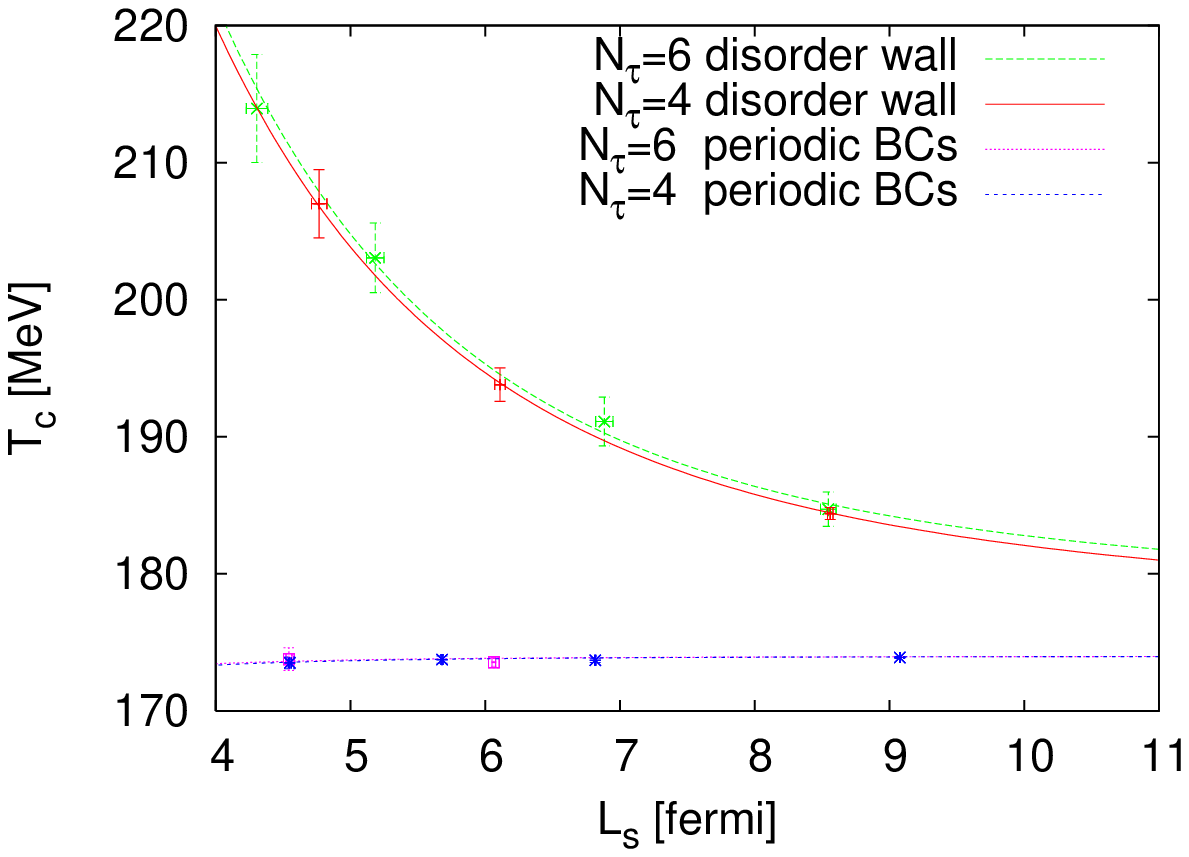,width=0.47\textwidth}
\caption{Fits of pseudo-transition coupling constant values (left). 
Estimate of finite volume corrections to $T_c$ (right). 
\label{fig_TcFits} }
\end{center} 
\end{figure}
The left Fig.~\ref{fig_TcFits} shows thus obtained fits of 
pseudo-transition coupling constant values versus $N_{\tau}/N_s$ 
(the $N_s\to\infty$ value is extrapolated from simulations with 
periodic BCs). Using the scaling relation ({\ref{LambdaLat}) we 
eliminate the coupling in favor of $T_c$ and $L_s$ and obtain the 
right Fig.~\ref{fig_TcFits}. There are {\it no free parameters} in 
this step, because the scaling relation was determined previously 
in independent work. The $N_{\tau}=4$ and $N_{\tau}=6$ data collapse
to one curve, i.e., despite the small values of the temporal lattice
sizes the results are perfectly consistent with scaling.

\begin{figure}[-t] \begin{center} 
\epsfig{figure=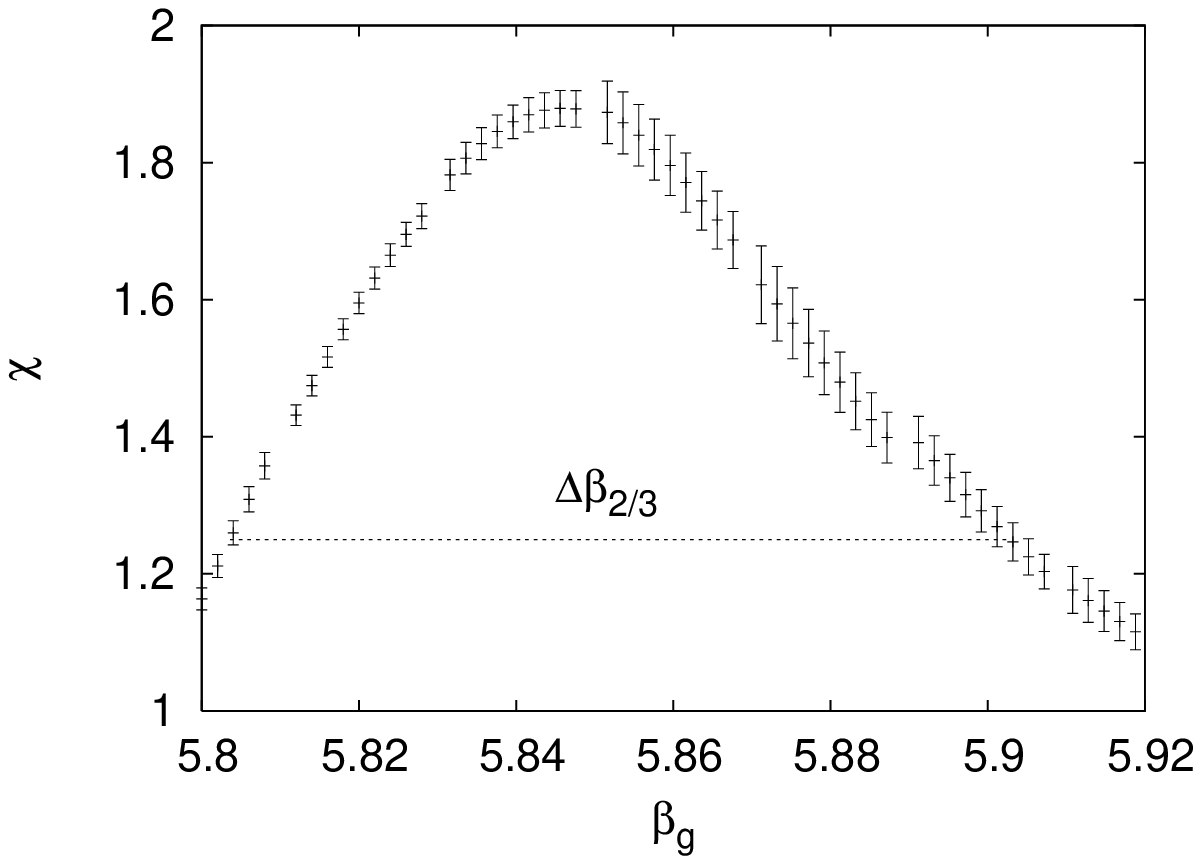,width=0.47\textwidth}
\hfill
\epsfig{figure=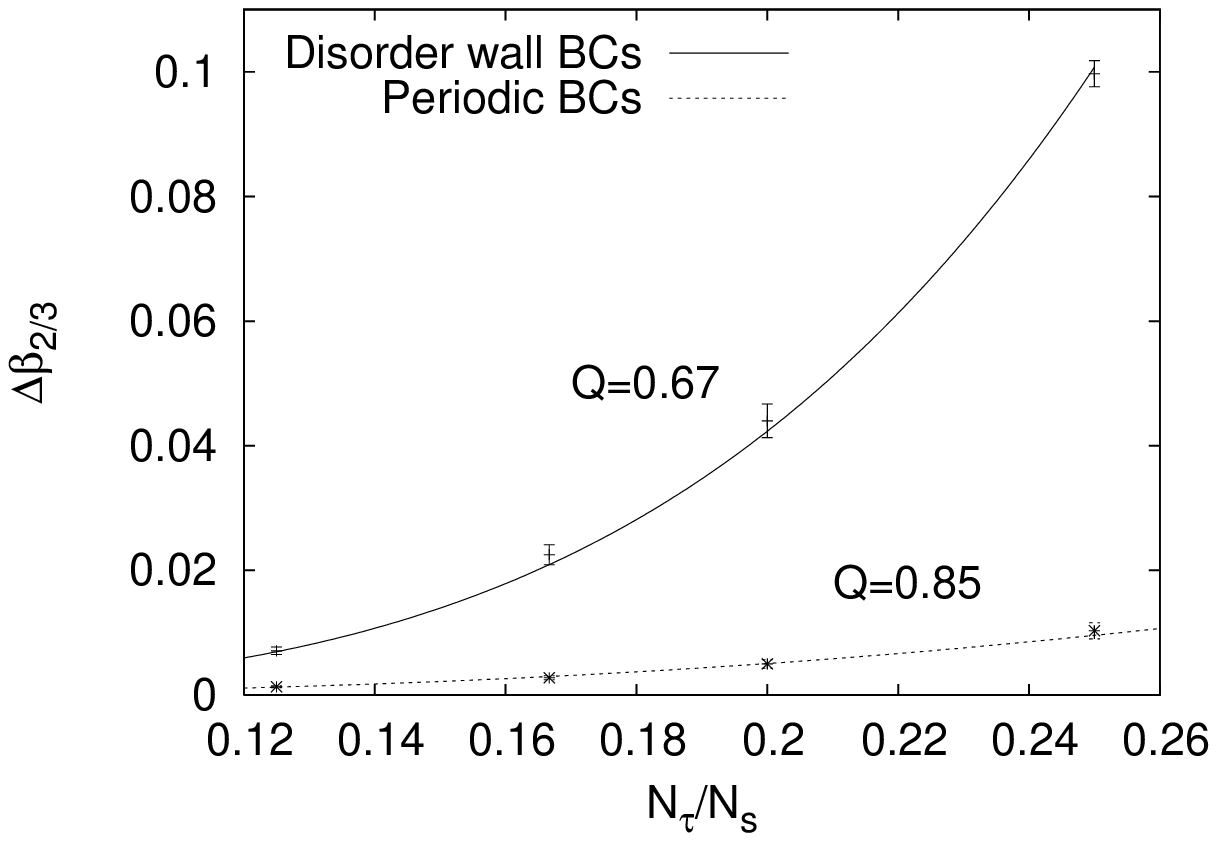,width=0.47\textwidth}
\caption{Polyakov loop susceptibility with disorder wall BCs 
on a $4\times 16^3$ lattice (left). Fits of the $N_{\tau}=4$ 
widths (right). \label{fig_width} }
\end{center} 
\end{figure}

The left Fig.~\ref{fig_width} shows the Polyakov loop susceptiblity
on a $4\times 16^4$ lattice with disorder BCs and its full width at 
2/3 maximum, which we used instead of the more conventional full 
width at half maximum, because the former is easier to extract from 
MC data (smaller reweighting range). Our width data are fitted to 
the form
\begin{equation}
  \Delta\beta^g_{2/3}=c_1^p\left(\frac{N_\tau}{N_s}\right)^3
                   +c_2^p\left(\frac{N_\tau}{N_s}\right)^6
\end{equation}
for periodic BCs and to
\begin{equation}
  \Delta\beta^g_{2/3}=c_1^d\left(\frac{N_\tau}{N_s}\right)^3
                   +c_2^d\left(\frac{N_\tau}{N_s}\right)^4
\end{equation}
for disorder wall BCs. The first term reflects in both cases the
delta function singularity of a first order phase transition, i.e.,
the width times the Polyakov loop maximum is supposed to approach a
constant for $N_s\to\infty$. The leading order correction to 
that is {1/Volume} for periodic BCs and $1/N_s$ for disorder 
wall BCs. Plots of the corresponding fits are shown in 
Figs.~\ref{fig_width} (right) and~\ref{fig_widthLs} (left).
As before, we use the scaling relation (\ref{LambdaLat}) to
eliminate the coupling constant and show in Fig.~\ref{fig_widthLs}
(right) the thus obtained volume dependence of the width of
the transition. Again, we see collapse to a nice scaling curve.

\begin{figure}[-t] \begin{center} 
\epsfig{figure=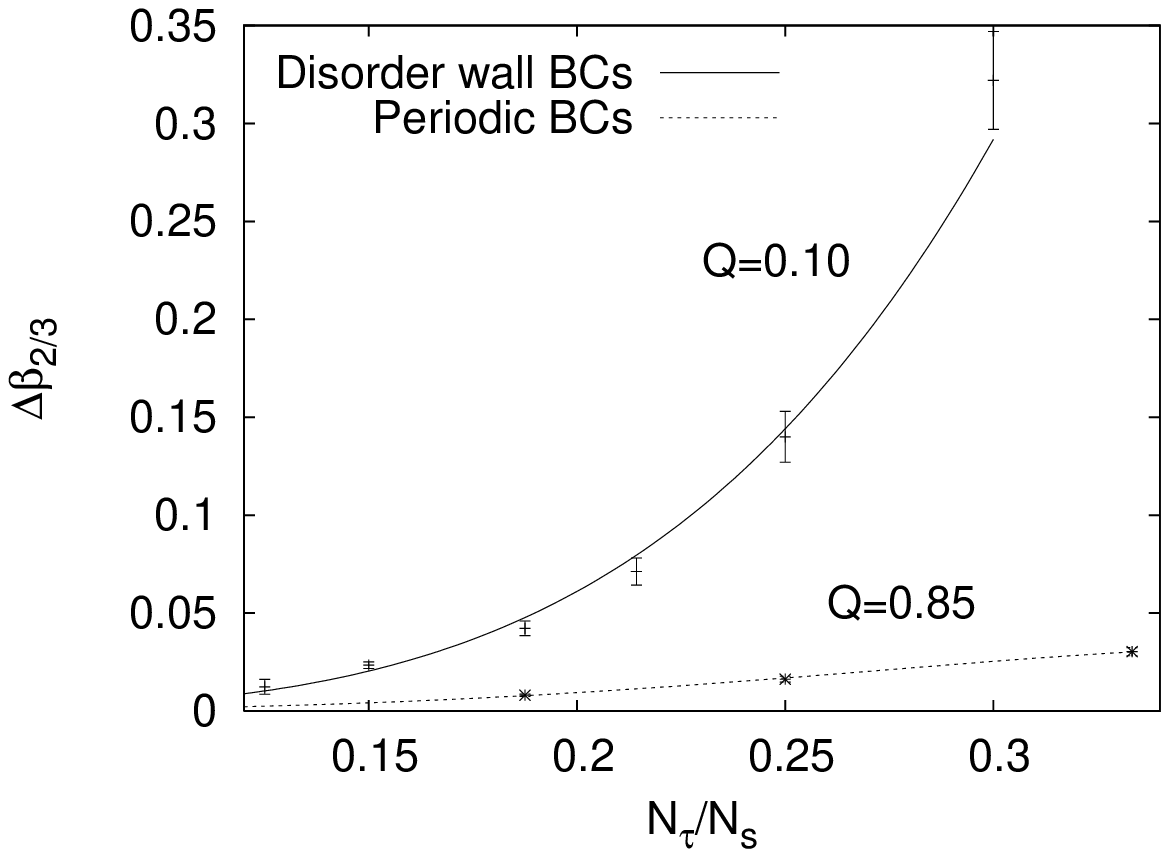,width=0.47\textwidth}
\hfill
\epsfig{figure=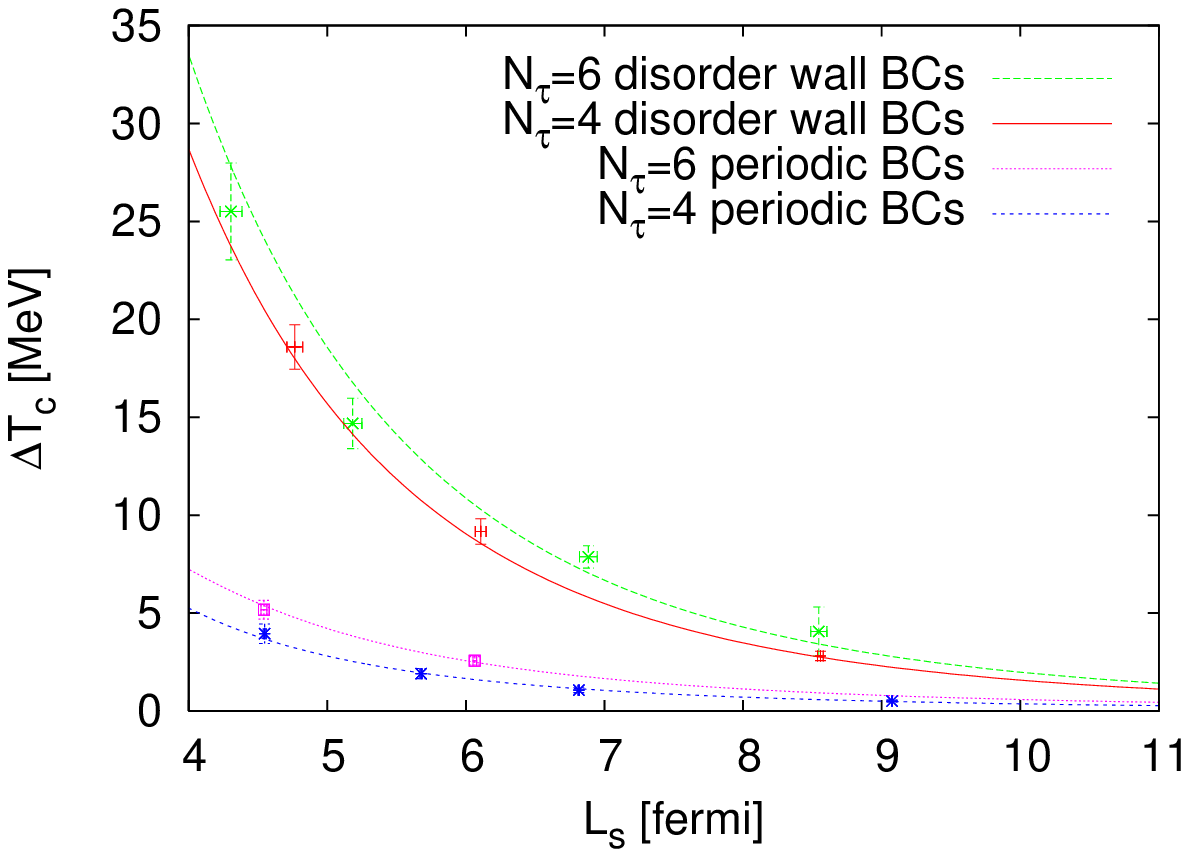,width=0.47\textwidth}
\caption{Fits of the $N_{\tau}=6$ widths (left). Estimate of finite 
volume corrections to the width (right).  \label{fig_widthLs} }
\end{center} 
\end{figure}

\section{Shortcomings of the Disorder Wall BCs}

The spatial lattice spacing $a_s$ should be the same on both sides of 
the boundary, but for the disorder wall this is not true. It reflects
the temperature jump at the price of introducing a similar jump in
the spatial lattice spacing. Its main advantage that it allows for
technically simple simulations, and one can hope that the temperature
jump is the only relevant quantity for the questions asked.

A construction, called {\it confinement wall} in \cite{BaBe07}, for 
which the physical length of one spacelike lattice spacing stays 
constant across the boundary can be achieved by using an anisotropic 
lattice for the volume $V_1$: 
\begin{equation} 
  S(\{U\}) = \frac{\beta^g_s}{3} \sum_{{\Box}_s} {\rm Re}\,
             {\rm Tr} \left( U_{{\Box}_s}\right)
           + \frac{\beta^g_{\tau}}{3} \sum_{{\Box}_{\tau}} {\rm Re}\,
             {\rm Tr} \left( U_{{\Box}_{\tau}}\right)\,.
\end{equation}
The lambda scale of this action has been investigated by Karsch 
\cite{Ka82} and in the continuum limit one finds
\begin{equation} 
  \beta^g_{\tau}/\beta^g_s = \left(a_s/a_{\tau}\right)^2\,.
\end{equation}
When we aim at $a_0=a_s\approx 10^{-10}\,a_{\tau}$ the sublattice 
$V_1$ is driven to $\beta^g_{\tau}=0$ and the simulation of the 
confined world becomes effectively 3D. However, in a first step one 
may be content with a temperature slightly below $T_c$ on the 
outside, so that the confinement wall allows to have all 
$\beta$-values in their scaling regions. Another approach may
want to rely on symmetric lattices to model low temperatures.

\section{Summary and Conclusions} 

\begin{enumerate} 
\item As noted before \cite{Bo96} finite size corrections to deconfinement
      properties of SU(3) are very small for periodic BCs.
\item For volumes of BNL RHIC size the magnitudes of SU(3) corrections 
      due to cold boundaries appear to be comparable to those of 
      including quarks into pure SU(3) LGT. Our data show the correct 
      SU(3) scaling behavior.
\item Extension of measurements should be done, to calculate
      the equation of state.
\item Previous calculations \cite{F07,K07} of full QCD at finite 
      temperatures and RHIC (low) densities should be extended to 
      other than periodic BCs.
\item There appears to be a variety of options to include cold 
      boundaries and approaching the finite volume continuum limit. 
      Therefore, more experience with pure SU(3) LGT is desirable 
      before including quarks. Next, we intend to focus on the 
      confinement wall with both couplings in the scaling region 
      (i.e., an outside temperature just below $T_c$).
\end{enumerate}

\acknowledgments We thank Urs Heller for discussions on 
the question of using symmetric lattices to model cold boundaries.
This work was supported by the US Department of Energy 
under contract DE-FG02-97ER41022.

\end{document}